\documentclass[12pt,a4paper]{article}

\advance\voffset by -2.0cm \advance\hoffset by -1.25cm
\def\AEF{A.E. Faraggi}

\def\NPB#1#2#3{{ Nucl.\ Phys.}\/ {\bf B#1} (#2) #3}
\def\PLB#1#2#3{{ Phys.\ Lett.}\/ {\bf B#1} (#2) #3}
\def\PLA#1#2#3{{ Phys.\ Lett.}\/ {\bf A#1} (#2) #3}
\def\PRD#1#2#3{{ Phys.\ Rev.}\/ {\bf D#1} (#2) #3}
\def\PRL#1#2#3{{ Phys.\ Rev.\ Lett.}\/ {\bf #1} (#2) #3}
\def\PRT#1#2#3{{ Phys.\ Rep.}\/ {\bf#1} (#2) #3}

\def\IJMP#1#2#3{{ Int.\ J.\ Mod.\ Phys.}\/ {\bf A#1} (#2) #3}

\def\JCAP#1#2#3{{ JCAP}\/ {\bf #1} (#2) #3}

\usepackage{amsmath}
\usepackage{amssymb}
\usepackage{amsthm}
\usepackage{amsfonts}
\theoremstyle{definition}

\newcommand{\RR}{\mathbb{R}} % Reali
\newcommand{\NN}{\mathbb{N}} % Naturali
\newcommand{\Sc}{\mathcal{S}} % ???????
\newcommand{\be}{\begin{equation}}
\newcommand{\ee}{\end{equation}}
\newcommand{\beqa}{\begin{eqnarray}}
\newcommand{\beqn}{\begin{eqnarray}}
\newcommand{\eeqn}{\end{eqnarray}}
\newcommand{\eeqa}{\end{eqnarray}}

\def\1{\frak 1}
\def\2{\frak 2}
\def\3{\frak 3}

\newlength{\oldcolsep}

\begin{document}

%\rightline{LTH-961}

\title{Energy Quantisation and Time Parameterisation}

\author{Alon E. Faraggi$^1$ and Marco Matone$^2$}\date{}

\maketitle

\begin{center}{$^1$Department of Mathematical Sciences\\
University of Liverpool,
Liverpool L69 7ZL, UK }\end{center}

\begin{center}$^2$Dipartimento di Fisica e Astronomia ``G. Galilei'' and Istituto
Nazionale di Fisica Nucleare \\
Universit\`a di Padova, Via Marzolo, 8 -- 35131 Padova,
Italy\end{center}

\bigskip

\bigskip

\begin{abstract}
We show that if space is compact, then trajectories cannot be defined in the framework of quantum Hamilton--Jacobi equation.
The starting point is the simple observation that when the energy is quantized it is not possible to make variations with respect to the energy,
and the time parameterisation $t-t_0=\partial_E{\Sc}_0$,
implied by Jacobi's theorem and that leads to group velocity, is ill defined.
It should be stressed that this follows directly form the quantum HJ equation without
any axiomatic assumption concerning the standard formulation of quantum mechanics.
This provides a stringent connection between the quantum HJ equation and the
Copenhagen interpretation. Together with tunneling and the
energy quantization theorem for confining potentials, formulated in the framework of quantum
HJ equation, it leads to the main features of the axioms of quantum
mechanics from a unique geometrical principle. Similarly to the case
of the classical HJ equation, this fixes its quantum analog by requiring
that there exist point transformations, rather than canonical ones, leading
to the trivial hamiltonian. This is equivalent to a basic cocycle
condition on the states.
Such a cocycle condition can be implemented on compact spaces, so
that continuous energy spectra are allowed only as a limiting case.
Remarkably, a compact space would also imply that the Dirac and von Neumann
formulations of quantum mechanics essentially coincide.
We suggest that there is
a definition of time parameterisation leading to trajectories in the context of the quantum HJ equation having
the probabilistic interpretation of the Copenhagen School.
\end{abstract}

\noindent

\newpage
\setcounter{footnote}{0}
\renewcommand{\thefootnote}{\arabic{footnote}}

\section{Introduction}

The validity of quantum mechanics is indisputable,
but leaves many conceptual problems unresolved.
For this reason over the years numerous schemes
have been proposed to address such issues. Among those,
the quantum HJ theory is one of the most investigated topics
(see \cite{QHJT} for a partial list of papers).
These studies involve the foundation of quantum mechanics,
in particular its interpretation, cosmology, the analysis of
quantum dynamics, molecular trajectories, etc. It has been
suggested that the Quantum Hamilton--Jacobi Equation (QHJE) yields a trajectory
representation of quantum mechanics (see e.g. \cite{poirier} and references therein).
The absence of trajectories is, however,
inherent in the Copenhagen probabilistic interpretation of the
quantum mechanical wave function.
There appears, therefore, to be a fundamental dichotomy between
the two approaches.
In this paper we offer a solution to this puzzle.
We demonstrate the absence of trajectories
in the derivation of the QHJE from point transformations leading to the trivial hamiltonian \cite{fm2}\cite{bfm}.
The basic point is that trajectories can
only be defined by time pamaterisation of them, and
include the Bohm--de Broglie pilot wave
representation and Floyd's time parameterisation \cite{floyd} by using
Jacobi theorem. We show
in this paper that these time parameterisations are ill
defined.
This resolves the dichotomy with the Copenhangen interpretation.

In \cite{fm1} we considered a Legendre duality in the framework of the
Schr\"odinger equation.
Consistency of this mathematical structure in the
context of the phase--space reveals that classical mechanics
must be modified \cite{fm2}.
The main idea has been to implement, as in the derivation of the classical
HJ equation,
the transformations leading to the free state. The difference is that we
did not consider the usual canonical transformation with the space
coordinate $q$ and the conjugate momentum $p$ considered as independent
variables. Rather,
we performed the transformation on $q$ and considered the transformation
on $p$ as the one induced by the coordinate transformation,
by assuming that the analog of the Hamilton characteristic function
transforms as a scalar field.
Such a transformation leads to a basic cocycle condition which, in turn,
implies that the analog of the Hamilton characteristic function must
satisfy the quantum analog of the stationary HJ equation.
The derivation extends to the higher dimensional non--stationary case and
to the relativistic case as well \cite{bfm}.

A basic theorem in \cite{fm2} is that
energy quantisation for bound states follows as a consistency condition.
We stress that whereas in the standard approach energy quantisation is a
direct consequence of the wave--function
interpretation, here it follows from the cocycle condition without any
assumption on the meaning of the wave--function.

The quantum Hamilton-Jacobi equation is
\begin{equation}
{\partial {\cal S}\over \partial t}+{1\over2m}(\nabla{\cal S})^2+V-{\hbar^2\over2m}{\Delta R\over R}=0 \ ,
\label{launo}\end{equation}
\begin{equation}
{\partial R^2\over\partial t}+{1\over m}\nabla\cdot (\nabla R^2{\cal S})=0 \ ,
\label{ladue}\end{equation}
which can be obtained by identifying a solution of the Schr\"odinger equation with $Re^{{i\over\hbar}{\cal S}}$. Note that ${\cal S}$ is the quantum analog of Hamilton's principal function.
Let us now consider a particle in a stationary state with energy $E$. We have
\begin{equation}
{\cal S}=\int p\cdot dq -E(t-t_0) \ ,
\label{sessezero}\end{equation}
where the first term, that we denote by ${\cal S}_0$, is the quantum analog of Hamilton's characteristic function.
Let now assume that such a particle admits a trajectory $q=q(t)$.
Performing a variation of time and energy, keeping fixed the initial and final spatial coordinates of the trajectory,
we have
$$
\delta {\cal S}={\delta{\cal S}_0\over\delta E}\delta E-(t-t_0)\delta E - E\delta t \ ,
$$
and by (\ref{sessezero})
\be
t-t_0={\partial{\Sc}_0\over\partial E} \ .
\label{unooo}\ee
This is the time parametrization of particle trajectories that, as first observed by Floyd \cite{floyd}, should be used
in considering the quantum HJ equation.
This is just how
trajectories are defined in classical mechanics as
it implies the group velocity.

A simple but basic initial observation is that in the case of quantized spectra it is not possible to make the variation of the energy.  In particular,
the trajectories $q=q(t)$, that would follow by inverting Eq.(\ref{unooo}), do not exist
in the case of discrete energy spectra. Consistency arguments also show
that this
excludes the possibility of using (\ref{unooo}) even in the case of continuous
energy spectra. It is clear that the non existence of trajectories suggests that also the
probabilistic interpretation may be derived without imposing it as a basic axiom.

As we will see, there is a connection between the non existence of trajectories and the cocycle condition.
In particular, we will see that the implementation of the cocycle condition
fixes gluing conditions on the ratio of two linearly independent solutions of
the wave--function
that can be satisfied only on compact spaces, so that continuous energy spectra arise only in the decompactification limit.
The fact that
such gluing conditions  are a basic step in the geometrical formulation
of \cite{fm2}\cite{bfm},
related to Legendre duality \cite{fm1}, and that led to the quantisation of
the energy in the case of confining potentials, suggests that they should be
always satisfied. It should be observed
that in the case of a free particle the level spacing of the energy spectrum
will be determined by the geometry of our three--dimensional space,
essentially of the order $R^{-2}$, where
$R$ is some characteristic cosmological length, and therefore extremely tiny.
This, of course, does not mean that such a spacing could not be detected a
priori. The fact that trajectories cannot be well--defined provides
an intriguing relation between quantum HJ theory and the Copenhagen
interpretation.
However, the reason why they do not exist in the quantum HJ theory is
that it is just the time parameterisation that does not exist.
In some sense a similar view is also the one of the Copenhagen
interpretation. Actually, time is not a (self--adjoint) operator,
so that also in standard quantum mechanics time is not an observable.

There are additional reasons to consider the possibility that a free particle
may have a quantized energy spectrum, even if the level spacing is
extremely tiny.
A rigorous treatment of continuous spectra requires elaborated structures
and the probability of finding the particle in a given volume cannot be
defined. Furthermore, it should be observed that there are two formulations
of quantum mechanics in the framework of the Copenhagen interpretation.
The one by von Neumann  \cite{vonNeumann},
where quantum states are always rays in the Hilbert space $L^2(\RR)$,
and the one by Dirac \cite{Dirac} that in general requires the rigged
Hilbert space and the eigenfunctionals. In particular,
in the Dirac formulation the wave--function of a free particle in $\RR^3$
is seen as a tempered distribution, that is an element in ${\Sc}'(\RR^n)$,
the dual space of the Schwarz space ${\Sc}(\RR^n)$.
In the case of compact spaces the Helmholtz equation
$$
-{\hbar^2\over2m}\Delta \psi=E\psi \ ,
$$
has only a discrete spectrum, so that the two formulations would essentially
coincide.
More generally, a compact space would fix some natural cutoff that may
play an important r\^ole.
We also consider the problem of defining a sort of probabilistic time
such that the resulting trajectories in the quantum HJ theory reproduce the
probabilistic interpretation
of the Copenhagen School.

\section{The cocycle condition and energy quantization}

We define
$${\cal W}(q)=V(q)-E \ ,
$$
 where $V$ is the potential energy and $E$ the energy
level.
The main point of \cite{fm2} is to assume the existence
transformations leading
any system to the one with trivial Hamiltonian, that is with ${\cal W}(q)=0$.
Doing this in classical mechanics leads to the classical HJ equation.
The difference in the approach of \cite{fm2} is that whereas in classical mechanics one considers the
canonical transformations, where $p$ and $q$ are considered independent, we considered point transformations. More precisely, we considered
the transformation of $q$ and fixed the transformation on the conjugate
momentum as the one
induced by the relation $p=\partial_q\Sc_0$, by considering $\Sc_0$,
the analog of the Hamilton characteristic function, as a scalar function under such a transformation of $q$.
The derivation can be extended to the
non--stationary case and to the relativistic version as well \cite{bfm}.
Therefore, the derivation of \cite{fm2,bfm} is based on the principle
that all physical state labeled by the function
${\cal W}(q)$ can be connected by a coordinate
transformation,
$$
q^a\rightarrow q^b=q^b(q^a) \ ,
$$ defined
by
$$
{\Sc}_0^b(q^b)={\Sc}_0^a(q^a) \ .
$$
This implies that
there always exists a coordinate transformation connecting
any physical state to the one with ${\cal W}^0(q^0)=0$. Inversely, this means
that any physical state can be reached from the
one with ${\cal W}^0(q^0)=0$ by a coordinate transformation.
This cannot be consistent
with Classical Mechanics (CM). The reason being that in
CM the physical system with ${\cal W}^0(q^0)=0$ remains a fixed
point under coordinate transformations. Thus, in CM it
is not possible to generate all systems by a coordinate
transformation from the trivial one. This
implies the modification of CM,
which is analyzed by a adding a still unknown function
$Q(q)$ to the classical HJ equation.
Consistency conditions then fix the
transformation properties for
${\cal W}(q)$,
$${\cal W}^v(q^v)=
 \left(\partial_{q^v}q^a\right)^2{\cal W}^a(q^a)+(q^a;q^v) \ ,
 $$
and
$$
Q^v(q^v)=\left(\partial_{q^v}q^a\right)^2Q^a(q^a)-(q^a;q^v) \ ,
$$
which fixes the cocycle condition
\be
(q^a;q^c)=\left(\partial_{q^c}q^b\right)^2[(q^a;q^b)-(q^c;q^b)] \ .
\label{cocycle}
\ee
The cocycle condition is invariant under M\"obius transformations
and fixes the functional form of the inhomogeneous term. Furthermore,
the cocycle condition fixes the identification
$${\cal W}(q)=-{\hbar^2\over{4m}}\{{\rm e}^{(2i{\Sc}_0/\hbar)},q\} \ ,$$
and
$${Q}(q)=  {\hbar^2\over{4m}}\{{\Sc}_0,q\} \ ,$$
where $\{f,q\}=f'''/f'-{3\over2}(f''/f')^2$ denotes the Schwarzian derivative.
The cocycle condition, that generalizes to higher dimensions \cite{bfm},
implies that
${\Sc}_0$ is solution of the Quantum Stationary
HJ Equation (QSHJE),
\begin{equation}
{1\over{2m}}\left({{\partial_q {\Sc}}_0}\right)^2+
V(q)-E+{\hbar^2\over{4m}}\{{\Sc}_0,q\}=0 \ .
\label{pethooft}\end{equation}
The equivalence with the one-dimensional stationary version of Eqs. (\ref{launo}) and (\ref{ladue}) follows by observing that by
Eq. (\ref{ladue}) one gets
$$
R=c {1\over \sqrt {\Sc_0'}} \ ,
$$
with $c$ a non--zero constant,
so that
$$
{\Delta R\over R}=-{1\over 2}\{{\Sc}_0,q\} \ .
$$
It is easy to check that the solution of (\ref{pethooft}) can be expressed in terms of solutions of the Schr\"odinger equation. In particular,
$${\rm e}^{{2i\over \hbar}{\Sc}_0}={\rm e}^{i\alpha}
{{w+i{\bar\ell}}\over {w-i\ell}} \ ,$$
where $\ell= \ell_1+i\ell_2$, with $\ell_1$ and $\ell_2\neq0$ two arbitrary constant, and
$w=\psi^D/\psi$, with $\psi^D$ and $\psi$
two real linearly independent solutions
of the Schr\"odinger equation
$$
\left(-{\hbar^2\over {2m}} {\partial^2\over
{\partial q}^2} + V(q) - E \right)\psi(q)=0 \ .
$$
A distinguished feature of
the formalism in \cite{fm2} is that both
solutions of the Schr\"odinger equation, $\psi$ and $\psi^D$,
are kept in the formalism. This can be seen
from the properties of the Schwarzian derivative that show that
the trivialising transformation is
$$
q\rightarrow q^0 =\gamma(\psi^D/\psi) \ ,
$$
where $\gamma(\psi^D/\psi)$ is an arbitrary M\"obius transformation of
$\psi^D/\psi$.
In general the wave--function in the
formulation of \cite{fm2} is
\be
\psi(q) =
R(q)\left(A {\rm e}^{{i\over\hbar}{\Sc}_0} + B {\rm e}^{-{i\over\hbar}
{\Sc}_0}\right) \ .
\label{ingenerals}\ee
Furthermore, consistency conditions imply that
${\Sc}_0(q)$ is never a constant. In particular,
the quantum potential $Q(q)$ is never trivial and plays the
r\^ole of intrinsic energy.

The formulation in \cite{fm2} extends to higher dimensions
 and to the relativistic case  as well \cite{bfm}.
Let us now review how energy quantisation arises
in our formalism.
The QSHJE is equivalent to the equation
$$\{w,q\}=-4m(V(q)-E)/\hbar^2 \ .
$$
This implies that $w\ne const$,  $w\in C^2(R)$ and
$w^{\prime\prime}$ differentiable on $\RR$.
In addition from the properties of the Schwarzian derivative it
follows that
$$
\{w,q^{-1}\}=q^4\{w,q\} \ ,
$$
which can be seen as a direct consequence of the cocycle condition.
However, such a relation is defined only if the conditions on
$w$ hold on the extended real line ${\hat \RR}=\RR\cup\{\infty\}$. That is
$w\ne const$, $w\in C^2({\hat \RR})$ with
$w^{\prime\prime}$ differentiable on $\RR$
and
\be
w(-\infty)=\left\{
\begin{array}{ll}
+w(+\infty) \, & {\rm if} \,\,\, w(-\infty)\ne \pm \infty \ ,\\
-w(+\infty) \, & {\rm if} \,\,\, w(-\infty)=\pm\infty \ .
\end{array}\right.
\label{winfty}
\ee
This means that $w$, and therefore the trivialising map, is a local
homeomorphism of $\hat {\RR}$ into itself. This implies the continuity of $(\psi^D,\psi)$ and
$(\psi^{D^\prime},\psi^\prime)$ without assuming the probability interpretation
of the wave--function.  In
particular, the QSHJE is defined
only if the ratio $w=\psi^D/\psi$ of a pair of real
linearly independent solutions of the Schr\"odinger equation
is a local homeomorphism of the extended real line
$\hat\RR=\RR\cup\{\infty\}$ into
itself. This is an important feature as the
$L^2(\RR)$ condition, which in the
Copenhagen formulation is a consequence
of the axiomatic interpretation of the
wave--function, directly follows as a basic theorem
which only uses the geometrical
gluing conditions of $w$ at $q=\pm\infty$.
In particular, denoting
by $q_-$ ($q_+$) the lowest (highest) $q$ for
which $V(q)-E$ changes sign, we have
\cite{fm2}

\bigskip

\noindent
{\it If}
\be
V(q)-E\geq\left\{\begin{array}{ll}P_-^2 >0 \ , & q<q_- \ ,\\ P_+^2 >0 \ , & q>
q_+ \ ,\end{array}\right.
\label{boundingpotential}\ee
{\it then $w=\psi^D/\psi$ is a local self--homeomorphism of
$\hat\RR$ iff the
Schr\"odinger equation has an
$L^2(\RR)$ solution.}

\bigskip

\noindent
Thus, since the QSHJE is defined if and only if $w$ is a
local self--homeomorphism of $\hat\RR$, this theorem implies that
energy quantisation {\it directly} follows from the
QSHJE itself
without further assumptions.

\section{Time parameterisation}

We emphasize that the present approach is fundamentally
distinct from the Bohmian one \cite{fm2}. Bohmian mechanics sets
$$\psi(q) = R(q) {\rm e}^{i S/\hbar} \ ,$$
where $\psi$ is the wave--function.
On the other hand, implementation of the point transformations leading to the trivial hamiltonian necessitates that the wave--function is
taken in the general form (\ref{ingenerals}). Such a condition is
reminiscent of the necessity in quantum field theories of
using the two solutions of the relativistic quantum equations.

In Bohmian mechanics
time parameterisation is defined by identifying $p$ with the
mechanical momentum
\be
p={{\partial {\Sc}}\over{\partial q}}= m {\dot q} \ ,\label{mechanicaltime}
\ee
with ${\Sc}$ solution of the quantum HJ equation.
In classical HJ theory time parameterisation is given by
\be
t-t_0={{\partial {\Sc}}_0^{\rm cl}\over{\partial E}} \ ,
\label{jacobitheorem}\ee
which leads to group velocity.
In Classical Mechanics (CM) this is equivalent to identifying the conjugate
and mechanical momenta. Namely, setting
$$p={\partial{\Sc}_0^{\rm cl}\over\partial q}=m{\dot q} \ ,$$
yields (\ref{jacobitheorem}).
Bohmian mechanics therefore brings back the notion of trajectories
for point particles, since we may solve for $q(t)$.
However, we note that the agreement between the definition
of time parameterisation of trajectories by (\ref{mechanicaltime})
and its definition by Jacobi theorem (\ref{jacobitheorem})
is no longer true in quantum mechanics. The use of the latter
definition in quantum mechanics, that is
\be
t-t_0= {{\partial {\Sc}_0}\over{\partial E}} \ ,
\label{floydproposal}
\ee
has been first proposed by Floyd \cite{floyd}.
In quantum HJ theory
this leads to
$$m{{dq}\over {dt}}=
{{\partial_q {\cal S}_0}\over{1-\partial_E(V+Q)}} \ .$$
Therefore, in quantum mechanics the time parameterisation in
(\ref{mechanicaltime})
does not coincide with its definition via Jacobi theorem.

Floyd proposal (\ref{floydproposal}) would in principle provide a
trajectory representation
of quantum mechanics which would
seem to be in contradiction with the inherently probabilistic nature
of quantum mechanics. We show that the definition
(\ref{floydproposal}) cannot be implemented. We emphasize, however, that this does not
mean that the parameterisation provided by (\ref{floydproposal})
cannot be useful. In fact, it is quite effective  to perform semi--classical
approximations. In this respect we note the successful application
in studies of molecular dynamics \cite{poirier}.

It is clear that in the case of quantized energy spectra
(\ref{floydproposal}) is not defined. On the other hand,
the concept of trajectory should be a universal one,
so that one should consider a time parameterisation which is
consistent in both cases of discrete and continuous spectra.
This essentially would exclude (\ref{floydproposal})
also in the case of continuous spectra. There is however a more
stringent argument to show that (\ref{floydproposal})
cannot be used to define time parameterisation of trajectories. In particular,
we now show that the QSHJE is defined only in the case of discrete
spectrum with the case of continuous spectrum
arising only as a limiting case.
We saw that the cocycle condition led to the QSHJE written as a
Schwarzian equation. In particular,
we established that the gluing conditions should always hold.
Let us now consider the case of a continuous spectrum.
Without loss of generality, we may consider the case when for large
$q$ the potential is zero. For large $q$ two associated
real independent solutions of the Sch\"odinger equation are
$$\psi=\sin kq \ , \qquad \psi^D=\cos kq \ .$$
This means that for large $q$ we have
$$
w=\tan k q \ .
$$
On the other hand, whereas this function at finite distance is a local
homeomorphism, it is not the case when considering the entire real axis.
This can be also seen by considering a free particle in
an infinitely deep well and considering the periodicity conditions.
This is equivalent to consider the particle in $S^1$ and then sending
its radius $R$ to infinity. For any finite $R$ the ratio of two linearly
independent solutions
is a local homeomorphism of $S^1$ and the spectrum is discrete. In the
$R\to\infty$ limit the gluing condition is no longer under control.
In the case of the free particle in an
infinitely deep potential well
of width $L$, the energy levels are
$$
E_n={\hbar^2\pi^2\over 2mL^2}n^2 \ ,
$$
$n\in{\NN}_+$.
It follows that for any but finite value of $L$ the spectrum is discrete.
We then conclude that imposing the cocycle condition requires a discrete
spectrum which in turn implies that continuous variation of the energy
is not possible.
Equivalently, the time parameterisation by Jacobi
theorem is not well defined.
This can be generalized to the free particle in any space dimension leading
to the same conclusion. On the other hand, it is a general theorem that the
equation
$$-{\hbar^2\over2m}\Delta \psi=E\psi \ ,
$$
in bounded domains of $\RR^n$, therefore including e.g. the $(n-1)$--sphere
$S^{n-1}$, has only a discrete spectrum. At the level of the quantum HJ equation,
this is just a consequence of the
non--triviality of the quantum potential.

Therefore, without using any axiomatic interpretation of the wave--function, we have shown that trajectories cannot be derived from the quantum HJ equation.
In particular, the concept of localized
particle with a defined velocity does not exist.
Since trajectories do not exist in the formulation of \cite{fm2}, one may try to consider finite differences instead of derivatives
\begin{equation}
t_n={{\Sc}_0(E_{n+1})-{\Sc}_0(E_{n})\over E_{n+1}-E_n} \ .
\label{ecco}\end{equation}
This is a basic point since it leads to consider the superposition of
different energy eigenstates, which
is at the basis of the interference phenomena. In particular, whereas
time evolution of a hamiltonian eigenfunction corresponds to an overall
phase, so that describing the same ray vector in the Hilbert space,
the role of time is apparent, through interference, just when the physical
system is the superposition of at least two energy levels.
Eq.(\ref{ecco}) indicates that the interference phenomena are deeply
related to the concept of time.

Whereas Eq.(\ref{ecco}) sheds light on the origin of interference in the
Quantum Hamilton-Jacobi formulation, there is an interesting alternative
which is suggested by
Scrh\"odinger equation.
One may think that time parameterisation  may be defined as a sort of
random parameter whose distribution is
fixed by the properties of the distribution of the energy eigenvalues.
In other words, instead of considering finite difference, that gives
$t=t(E)$, we may search for a more intrinsic definition.
The idea is that the wave--function depends on $E$,
so one may formally invert this relation to
$E=E(\psi)$. This leads to consider
$t_\psi=t(\psi)$. In particular, consider a time independent potential. In this case the
Schr\"odinger equation
implies
\begin{equation}
\psi(q,t)=\sum_k c_k e^{-{i\over\hbar}E_kt}\psi_k(q) \ ,
\label{evoluzione}\end{equation}
where the $\psi_k$ are the hamiltonian eigenfunctions
$$
H\psi_k=E_n\psi_k \ .
$$
Next, recall that for one-dimensional bound states $\psi_n$ can be normalized in such a way that it takes real values.
Multiplying (\ref{evoluzione}) by $\psi_n$ and integrating over $\RR$ one gets
\be
t_\psi={i\hbar\over E_n}\Big(\ln \int_\RR \psi \psi_n dq -
\ln \int_\RR \psi(q,0)\psi_n dq\Big) \ .
\label{fondamentaleeee}\ee
Note however that $t_\psi$ is just the time experienced by the observer.
This suggests that there exists a kind of time
parameterisation leading to the probabilistic interpretation of
trajectories in the framework of the quantum HJ equation.

\section{Conclusions}\label{conclude}

Understanding the synthesis of quantum mechanics and general relativity
remains the pivotal goal of fundamental physics.
The main effort in this endeavour is in the framework of
string theory. String theory provides a self--consistent
perturbative approach to quantum gravity. The main achievement of
string theory is that it gives rise to the gauge and matter
ingredients of elementary particle physics, and predicts the
number of degrees of freedom required to obtain a consistent theory.
String theory therefore enables the development of a phenomenological
approach to quantum gravity. The state of the art in this regard is
the construction of Minimal Standard Heterotic String Models
\cite{mshsm}. Over the past few years important progress
has also been achieved in the understanding of the perturbative
expansion of string theory for genus $g$ Riemann surfaces, see for example \cite{marcogg} and references therein.
Despite its successes, string theory does not provide a conceptual starting
point for formulating quantum gravity. What we seek is a fundamental hypothesis
of which, possibly, string theories arise as perturbative limits.
A basic property of string theory is T--duality \cite{tdual},
which may be viewed as phase--space duality in compact space.

In this paper we examined
the definability of time parameterisation of trajectories in the framework of the quantum HJ equation. We first observed
that it is not possible to perform infinitesimal variations of the energy when this is quantized. As such, the quantum HJ equation does not lead to
the concept of trajectory. Next, we discussed the identification
the mechanical momentum with the conjugate momentum
$$
m\dot q={\partial{\cal S}\over\partial q} \ ,
$$
as done in Bohmian mechanics, and noticed that it is inconsistent
with its derivation from Jacobi's theorem, which in turn cannot be applied in the case of systems with quantized energy.
On the other hand, consistency arguments show that if trajectories cannot be defined in the case of of quantized energy, then trajectories do not exist even when
the energy spectrum is continuous.

In this respect it should be however stressed that quantum trajectories provide a powerful approximation tool in quantum dynamics
 \cite{poirier}\cite{Wyatt}. In particular, the method of quantum trajectories has been developed as a computational tool to solve time-dependent quantum mechanical problems by evolving ensembles of correlated quantum trajectories through the integration of the hydrodynamic equations. These quantum trajectories serve as a computational adaptive moving grid and from this perspective, these are not authentic trajectories.

We then proposed that the above features
of the quantum HJ equation are at the heart of the
probabilistic nature of quantum mechanics. In this respect, we considered time as dependent on the wave--function getting an expression that suggests considering time itself as having a
probabilistic nature.

We also observed that the derivation of the quantum HJ equation from the cocycle condition \cite{fm2} is naturally formulated in compact spaces that can be extended considering the
decompactification limit. This suggests that our space is compact. In this case all the possible energy spectra are quantized and would make essentially equivalent the Dirac and von Neumann formulations of quantum mechanics.
The compactness of space has basic observational consequences
and forthcoming evidence for it may exist in the
cosmic microwave background radiation \cite{aa}.

\section*{Acknowledgements}

AEF is supported by STFC under contract ST/J000493/1.
MM is supported by the Padova University Project CPDA119349
and by the MIUR-PRIN contract 2009-KHZKRX.

\newpage


\end{document}